\documentclass[12pt,dvips,twoside,letterpaper]{article}
\usepackage[usenames]{color}

 \usepackage[bookmarks=false]{hyperref}

\usepackage{pslatex}
\usepackage{fancyhdr}
\usepackage{graphicx}
\usepackage{geometry}

\usepackage{lineno,hyperref}
\modulolinenumbers[5]

\usepackage{amsmath}
\usepackage{amssymb}
\usepackage{amsfonts}
\usepackage{amsthm,amscd}

\newtheorem{theorem}{Theorem}

\theoremstyle{definition}
\newtheorem{definition}{Definition}

\theoremstyle{remark}

\newcommand{\eq}[1]{(\ref{#1})}

\def\C{\mathbb C}

\def\R{\mathbb R}

\def\cal{\mathcal}

\def\HH{\mathcal{H}}

\begin{document}

\centerline{\bf A minimalist approach to conceptualization of time
in quantum theory}

\bigskip

\centerline{Hitoshi Kitada$^a$, Jasmina Jekni\' c-Dugi\' c$^b$,
Momir Arsenijevi\' c$^c$, Miroljub Dugi\' c$^c$}

\bigskip

$^a${Department of Mathematics, Metasciences Academy, Japan}

$^b${Department of Physics, Faculty of Science and Mathematics,
University of Ni\v s,  Serbia }

$^c${Department of Physics, Faculty of Science, University of
Kragujevac,  Serbia }

\begin{abstract}

Ever since Schr\" odinger, Time in quantum theory is postulated
Newtonian for every reference frame. With mathematical rigor, we
show that the concept of the so-called Local Time allows avoiding
the postulate. In effect,  time appears as neither fundamental nor
universal on the quantum-mechanical level while being consistently
attributable to every, at least approximately, closed  quantum
system as well as to every of its (conservative or not)
subsystems.

\end{abstract}


\linenumbers

\section{Introduction}\label{Introduction}
Schr\"odinger's Quantum Mechanics in \cite{{S1}, {S2}, {S3}},
 is timeless when he introduced his fundamental equation
as a time-independent equation
\begin{equation}
H\psi=E \psi.\label{stationarySchrodingerEquation}
\end{equation}
Here $E\in \R$ and the Hamiltonian $H$ is of the form
\begin{equation}
H=\frac{\hbar^2}{2m}p^2+V(x),\quad
V(x)=-\frac{e^2}{|x|},\label{Hamiltonian}
\end{equation}
where
\begin{equation}
p=\frac{1}{i}\frac{\partial}{\partial
x}=\frac{1}{i}\left(\frac{\partial}{\partial
x_1},\frac{\partial}{\partial x_2},\frac{\partial}{\partial
x_3}\right)
\end{equation}
is the momentum operator conjugate to the position operator
$x=(x_1,x_2,x_3)$. With this stationary Schr\"odinger equation, he
could successfully give an explanation of the spectral structure
of hydrogen atoms, showing that his formulation of quantum
mechanics as the eigenvalue problem of a partial differential
operator is valid. Later he proved in \cite{SH} that his
formulation is equivalent with Heisenberg's formulation of QM.
Without loss of generality, we assume $m=1$ later on.

In the subsequent part \cite{S4} he emphasized the necessity to
give a time-dependent expression of the equation in order to treat
the nonconservative systems, and gave a time-dependent
 equation for general Hamiltonians
\begin{equation}
\frac{\hbar}{i}\frac{d\psi}{dt}(t)+H\psi(t)=0.\label{time-dependent-SchrodingerEquation}
\end{equation}
Schr\"odinger then applied the equation to some time-de\-pen\-dent
perturbations with an emphasis of the advantage of the
time-dependent approach. He however gave no justification for the
notion of time which is assumed for the equation. That is, ``time"
is postulated \cite{S4} to be unique and universally valid
throughout the universe as Newton put it in his {\it Principia
Mathematica}.

Exactly the same physical nature of time is assumed for the
standard text-book approach to quantum dynamics that is based on
the unitary operator $U(t)$, which defines a dynamical map for
 quantum systems, $\Psi(t) = U(t) \Psi(t=0)$. Hence we can
detect the following two assumptions (postulates) built in the
fundamental equation for quantum systems dynamics. The first
assumption is the equation's mathematical form provided by eq.(4),
which here we adopt without modification. The second assumption is
that quantum dynamics unfolds within the classical Newtonian
universal (global) time. However, at least as a logical
possibility, removing the second assumption is not excluded and,
if successful, might make the quantum foundations even more
efficient--the less number of postulates, the better theory.


Avoiding this assumption is not a trivial task, which we undertake
in this paper.
Rejecting the in-advance-agreed role of ``physical time'' for the
parameter $t$ in the unitary operator $U(t)$ elevates to the
following two related problems. First, if not in advance, then
certainly {\it a posteriori} the role of the parameter $t$ as
physical time should be rigorously established; non-rigorous
procedures typically assume certain additional rules and
assumptions, often of the interpretational relevance, that here we
are not interested in. Second,  without a postulate or an
interpretational framework, it is not obvious how to link the
time-independent Hamiltonian of closed system with the notion of
time. These subtle points are regarded with mathematical rigor in
Sections 4 and 5 with the general mathematical basis provided in
Section 2.
 As a result, in Sections 3 and 4 we emphasize a
possibility to introduce a notion of time for an arbitrary
 (including many-particle) {\it closed} system with the {\it time-independent} Hamiltonian.
We perform
  {\it  without}
 resorting to any {\it ad hoc} procedures or additional assumptions -- such as existence of the system's environment,
 be it classical \cite{BriggssRostEPJD} or not, or time quantization \cite{Y.AharonovD.Bohm}, or in-advance-agreed
 character of physical time. Expectably, such possibility comes at certain price, which in our approach is that time is
 neither fundamental nor universal on the quantum-mechanical level, and can be recognized as
 the so-called  (quantum-mechanical) local time \cite{K}.


\section{$N$-particle system}\label{N-particle}
In this section we consider a general {\it conservative} (i.e.
{\it closed}) quantum mechanical system consisting of $N$
particles; we take a unit system such that $\hbar = 1$. For such a
system of $N(\ge 2)$ quantum mechanical particles with mass
$m_i(>0)$ located at $r_i\in \R^3$ $(i=1,\dots,N)$,  Hamiltonian
\eq{Hamiltonian} becomes
\begin{equation}
\begin{aligned}
H=-\sum_{i=1}^N\frac{1}{2m_i}\Delta_{r_i}+V(x)=\sum_{i=1}^N\frac{1}{2m_i}\left(\frac{1}{i}\frac{\partial}{\partial
r_i}\right)^2+V(x),\\ V(x)=\sum_{1\le i< j\le
N}V_{ij}(x_{ij}),\label{N-bodyHamiltonian}
\end{aligned}
\end{equation}
where $\Delta=\Delta_{r_i}=\left(\frac{\partial}{\partial
r_i}\right)^2=\sum_{j=1}^3\frac{\partial^2}{\partial r_{ij}^2}$
$(r_i=(r_{i1},r_{i2},r_{i3})\in\R^3)$ is Laplacian and
$V_{ij}(x_{ij})$ $(x_{ij}=r_i-r_j)$ is a pair potential working
between the pair of particles $i$ and $j$. When we consider the
relative motion of $N$ particles, we can separate the motion of
the center of mass as follows. The center of mass of this
$N$-particle system is
\begin{equation}
X_C=\frac{m_1r_1+\dots+m_Nr_N}{m_1+\dots+m_N}.
\end{equation}
Defining the Jacobi coordinates by
\begin{equation}
\begin{aligned}
x_i=(x_{i1},x_{i2},x_{i3})=r_{i+1}-\frac{m_1r_1+\dots+m_ir_i}{m_1+\dots+m_i}(\in\R^3),\\(i=1,\dots,N-1)
\end{aligned}
\end{equation}
and corresponding conjugate momentum operators by
\begin{equation}
P_C=\frac{1}{i}\frac{\partial}{\partial X_C},\ \ \
p_i=\frac{1}{i}\frac{\partial}{\partial
x_i}=\frac{1}{i}\left(\frac{\partial}{\partial
x_{i1}},\frac{\partial}{\partial x_{i2}},\frac{\partial}{\partial
x_{i3}}\right),
\end{equation}
we decompose the Hilbert space $L^2(\R^{3N})$ as a tensor product
$L^2(\R^{3N})=L^2(\R^3)\otimes\HH$, $\HH=L^2(\R^{3n})$ with
$n=N-1$. Accordingly the Hamiltonian $H$ in \eq{N-bodyHamiltonian}
is decomposed as follows.
\begin{equation}
\begin{aligned}
&H=H_C\otimes I+I\otimes \tilde H,\\
&\tilde H=\tilde H_0+V,\quad H_C=\frac{1}{\sum_{j=1}^Nm_j}P_C^2,\\
&\tilde H_0=\sum_{i=1}^{N-1}\frac{1}{2\mu_i}p_i^2,\\
&\mu_i^{-1}=m_{i+1}^{-1}+(m_1+\dots+m_i)^{-1}\quad(i=1,\dots,n).
\end{aligned}\label{NHamiltonianseparated}
\end{equation}
Here $I$ denotes the identity operator. For real potentials
$V_{ij}(x_{ij})$, $H$ in \eq{N-bodyHamiltonian} and $\tilde H$ in
\eq{NHamiltonianseparated} define self-adjoint operators in the
Hilbert spaces $L^2(\R^{3N})$ and $\HH=L^2(\R^{3n})$,
respectively, and the relative motion of the $N$-particles is
described by the Hamiltonian $\tilde H$ in $\HH=L^2(\R^{3n})$.

By \eq{NHamiltonianseparated}, $H_C\otimes I$ is a nonnegative
selfadjoint operator in $L^2(\R^{3})$ and describes the free
motion of the center of mass of the $N$-particle system whose
property is well-known. Our main concern is thus about the
relative motion of the $N$ particles. Henceforth we will write
\begin{equation}
H=\tilde H, \quad H_0=\tilde H_0,
\end{equation}
and consider the Hamiltonian in $\HH=L^2(\R^{3n})$
\begin{equation}
H=H_0+V=\sum_{i=1}^{N-1}\frac{1}{2\mu_i}p_i^2+V(x).\label{HamiltonianofNbodysystem}
\end{equation}
We note that $H$ is defined solely through the configuration
operators $x=(x_1,\dots,x_{N-1})$ and conjugate momentum operators
$p=(p_1,\dots,p_{N-1})$. Thus time-independent QM is completely
determined through position and momentum operators $(x,p)$, since
the corresponding stationary time-independent Schr\"odinger
equation \eq{stationarySchrodingerEquation} is written as follows.
\begin{equation}
(H-\lambda I)\psi=0.\label{stationarySchrodingerEquation-1}
\end{equation}
This equation has non-zero solution $\psi\in\HH$ only when
$\lambda$ is an eigenvalue of $H$: $\lambda\in \sigma_p(H)$. A
complex number $\lambda$ is said to belong to the resolvent set
$\rho(H)$, when \eq{stationarySchrodingerEquation-1} has only a
trivial solution $f=0$ and the bounded inverse $(H-\lambda
I)^{-1}:\HH\rightarrow \HH$ exists.
$R(\lambda)=R_H(\lambda)=(H-\lambda I)^{-1}$ is called the
resolvent at $\lambda\in\rho(H)$ of $H$. We review some concepts
on spectrum $\sigma(H)$ of a selfadjoint operator $H$.
\begin{definition}\label{spectrum}
{
\begin{enumerate}
\item[ 1)]
 The set of all complex numbers $\lambda\in \C\setminus\rho(H)$ is called the spectrum of $H$ and denoted by $\sigma(H)$. For a selfadjoint operator $H$ it is trivial to see that $\sigma(H)\subset \R$.
\item[ 2)] We denote the resolution of the identity corresponding
to a selfadjoint operator $H$ by $E_H(\lambda)$ $(\lambda\in\R)$:
\begin{equation}
\begin{aligned}
&E_H(\lambda)E_H(\mu)=E_H(\min(\lambda,\mu)),\\
&\mbox{s-}\lim_{\lambda\to-\infty}E_H(\lambda)=0,\quad\mbox{s-}\lim_{\lambda\to\infty}E_H(\lambda)=I,\\
&E_H(\lambda+0)=E_H(\lambda),\\
&f(H)=\int_{-\infty}^\infty f(\lambda) dE_H(\lambda)\quad(\forall
f\in C(\R)),
\end{aligned}\label{resolutionofidentity}
\end{equation}
where
$E_H(\lambda+0)=\mbox{s-}\lim_{\mu\downarrow\lambda}E_H(\mu)$ and
$C(\R)$ is the set of all complex-valued continuous functions on
$\R$. An operator-valued measure $E_H(B)$ $(B\subset
\R:\mbox{Borel set})$ is defined by the relation
$E_H((a,b])=E_H(b)-E_H(a)$ for $-\infty<a<b<+\infty$. \item[ 3)]
Set $P(\lambda)=E_H(\lambda)-E_H(\lambda-0)$ $(\lambda\in \R)$. We
note that $P(\lambda)\ne0$ iff $\lambda$ is an eigenvalue of $H$.
When $\lambda\in \sigma_p(H)$, $P(\lambda)\HH$ is the eigenspace
of $H$ for $\lambda\in\sigma_p(H)$. The pure point spectral
subspace (or eigenspace) $\HH_p(H)$ for $H$ is defined as the
closed linear hull of the set
\begin{equation}
\bigcup_{\lambda\in\R}P(\lambda)\HH.
\end{equation}
Eigenprojection $P_H$ is the orthogonal projection onto
$\HH_p(H)$. \item[ 4)] The continuous spectral subspace for $H$ is
defined by
\begin{equation}
\begin{aligned}
\HH_c(H)=\{\psi\ |\ E_H(\lambda)\psi & \\ \mbox{ is continuous
with respect to }\lambda\in\R\},
\end{aligned}
\end{equation}
and the absolutely continuous spectral subspace for $H$ by
\begin{equation}
\begin{aligned}
\HH_{ac}(H)&=\{\psi\ |\ \mbox{The measure}
 (E_H(B)\psi,\psi)= \\&=
 \Vert
E_H(B)\Vert^2 \mbox{is absolutely continuous with } \\&
\mbox{respect to Lebesgue measure on }\R\}.
\end{aligned}
\end{equation}

The singular continuous spectral subspace $\HH_{sc}(H)$ is defined
by $\HH_{sc}(H)=\HH_{c}(H)\ominus\HH_{ac}(H)$. Then the relation
$\HH=\HH_p(H)\oplus\HH_c(H)=\HH_p(H)\oplus\HH_{ac}(H)\oplus\HH_{sc}(H)$
holds. \item[ 5)] The part $H_p, H_c, H_{ac}, H_{sc}$ of $H$ in
$\HH_p(H), \HH_{c}(H)$, $\HH_{ac}(H),\HH_{sc}(H)$ are called
spectrally discontinuous, spectrally continuous, spectrally
absolutely continuous and spectrally singular continuous,
respectively. The spectra $\sigma(H_p),\sigma(H_c),$
$\sigma(H_{ac})$, $\sigma(H_{sc})$ are called point spectrum,
continuous spectrum, absolutely continuous spectrum and singular
continuous spectrum of $H$, and denoted by $\sigma_p(H)$,
$\sigma_c(H)$, $\sigma_{ac}(H),\sigma_{sc}(H)$, respectively.
\end{enumerate}  }
\end{definition}
For rather general pair potentials $V_{ij}(x_{ij})$, it is known
(\cite{PSS}) that the singular continuous spectrum
$\sigma_{sc}(H)$ is absent: $\HH_{sc}(H)=\{0\}$. Therefore we
assume henceforth that $\HH_c(H)=\HH_{ac}(H)$ and
$\HH=\HH_{p}(H)\oplus\HH_{ac}(H)$ hold.

The resolution of the identity $\{E_H(\lambda)\}_{\lambda\in\R}$
gives the spectral property of the selfadjoint operator $H$, and
completely determines $H$ in time-independent manner. In this
sense it gives a stationary formulation of the QM system $(H,\HH)$
with the Hamiltonian $H$ in a Hilbert space $\HH$.

\section{A missing link in the Schr\" odinger's approach}\label{missinglink}

In order to illustrate the idea behind our approach to quantum
dynamics, let us return to Schr\"odinger's tho\-u\-ghts
nevertheless without historical rigor. Schr\"odinger \cite{S4}
starts with wave equation
\begin{equation}
\Delta\psi-\frac{2(E-V)}{E^2}\frac{\partial^2\psi}{\partial
t^2}=0.\label{(1)}
\end{equation}
As the energy factor $E$ suggests, he has been implicitly assuming
the relation (bearing in mind $\hbar = 1$)
\begin{equation}
\psi \sim \mbox{Re }\left(e^{\pm iEt}\right),\label{(2)}
\end{equation}
which he regarded equivalent to
\begin{equation}
\frac{\partial^2\psi}{\partial t^2}=-{E^2}\psi\label{(3)}
\end{equation}
or (if complex-valued wave function $\psi$ is permitted)
\begin{equation}
\frac{\partial\psi}{\partial t}=\pm{ i} E\psi.\label{(3')}
\end{equation}
From \eq{(1)} and \eq{(3)} one has time-independent Schr\"odinger
equation:
\begin{equation}
\left(-\frac{1}{2}\Delta+V-E\right)\psi=0.
\end{equation}
Substituting \eq{(3')} gives time-dependent equation
\begin{equation}
\frac{1}{i}\frac{\partial\psi}{\partial
t}\pm\left(-\frac{1}{2}\Delta+V\right)\psi=0.\label{timedependentSchrodinger}
\end{equation}
When $\psi$ satisfies one of the equations
\eq{timedependentSchrodinger}, the complex conjugate
$\overline{\psi}$ satisfies the other, so that one can adopt one
of the equations as time-dependent Schr\"odinger equation:
\begin{equation}
\frac{1}{i}\frac{\partial\psi}{\partial
t}+\left(-\frac{1}{2}\Delta+V\right)\psi=0.\label{timedependentSchrodinger2}
\end{equation}

Even if disregarding some curious points about the assumptions
\eq{(2)}-\eq{(3')}, we note that there is a large discrepancy
between the starting equation \eq{(1)} and the resulting equation
\eq{timedependentSchrodinger2}: \eq{(1)} is a wave equation and
the wave function $\psi(t)$ propagates with a constant velocity
$\sqrt{\frac{E^2}{2(E-V)}}$ if we ignore that it might be a
complex number. However, eq.\eq{timedependentSchrodinger2} is {\it
not} a wave equation and should, in turn, somehow describe also
the particle-aspect of quantum systems, in the sense of the
standard formalism based on the fundamental position (for brevity
denoted $x$) and momentum (denoted $p$) observables, which provide
the ultimate basis for defining the system's Hamiltonian. Thus
time-dependent Schr\"odinger equation
\eq{timedependentSchrodinger2} does not describe the wave function
propagating with constant velocity. Hence  a missing link in the
derivation of the fundamental equation
\eq{timedependentSchrodinger2} for {\it closed} systems.

As emphasized in Introduction, Time is generally tho\-u\-ght to be
unique and valid throughout the universe, and when we admit the
missing link between equations \eq{(1)} and
\eq{timedependentSchrodinger2}, one usually regards it a problem
of the choice of equation under a given universal time. In this
framework of thought, Schr\"odinger had chosen
\eq{timedependentSchrodinger2} without giving any justification
for the choice.

However if we see the problem closely, we will notice that we can
see it as the problem {\it which notion of time} we should choose.
For illustration let us suppose that $V=0$ and $E=2$ for the time
being. Then equations \eq{(1)} and \eq{timedependentSchrodinger2}
can be written respectively as follows.
\begin{align}
&\frac{1}{i}\frac{\partial\psi}{\partial t}+H^{(1)}\psi=0,\quad H^{(1)}=(-\Delta)^{1/2},\label{(1)'}\\
&\frac{1}{i}\frac{\partial\psi}{\partial t}+H^{(2)}\psi=0,\quad
H^{(2)}=-\frac{1}{2}\Delta.\label{(2)'}
\end{align}
Comparing \eq{(1)'} and \eq{(2)'}, we see that the rates of change
of the state $\psi$ with respect to the same change of time $t$
are different between the two equations. The rate for \eq{(1)'} is
\begin{equation}
H^{(1)}=(-\Delta)^{1/2}\label{H1}
\end{equation}
and that for \eq{(2)'} is
\begin{equation}
H^{(2)}=-\frac{1}{2}\Delta.\label{H2}
\end{equation}

However, we can also approach this from the following perspective.
We can assume that both equations \eq{(1)'} and \eq{(2)'} are {\it
correct} while describing {\it different processes} for the
systems that are subject to {\it different times}, which are
generated by the respective Hamiltonians \eq{H1} and \eq{H2}.
Hence the new role of the Hamiltonian: instead of solely
determining the rate of change of the system's state, the system's
Hamiltonian is recognized also to determine the time, which the
system is subjected to.

\section{Local time and clock, and a justification of the notion of time in quantum mechanics}\label{localtimeandlocalclock}

Generalizing the above argument, we will define the time of a
system with Hamiltonian $H$ in accord with \cite{K} as follows. We
will call $(H,\HH)$ a local system when a selfadjoint Hamiltonian
$H$ of a closed system is given in a Hilbert space $\HH$. Then we
can differentiate time among different systems, and the time which
is valid only for a {\it single closed} QM system $(H,\HH)$ will
be called the local time for the system.

To make the situation clear, we define local clock and local time
for a quantum mechanical system with Hamiltonian $H$ in
\eq{HamiltonianofNbodysystem} as follows. We note that $H$ is a
selfadjoint operator defined in a Hilbert space
$\HH=L^2(\R^{3n})$.

\begin{definition}\label{localtime-def}
The unitary group $e^{-itH}$ is called a {\it local clock} of the
local (closed) system $(H,\HH)$. The parameter $t$ in the exponent
of the local clock $e^{-itH}$ is called the {\it (quantum
mechanical) local time} for the system $(H,\HH)$.
\end{definition}

Essential in Definition \ref{localtime-def} is that it does {\it
not in advance} establish the physical meaning of the continuous,
real parameter $t$, which is dubbed ``local time''. Definition
\ref{localtime-def} only postulates the unitary dynamical map
$e^{-itH}$, which is generated by the closed system's Hamiltonian
with the necessarily appearing a $c$-number denoted $t$. Formally
it is clear that $\psi(t)=e^{-itH}\psi$ satisfies the
time-dependent Schr\"odinger equation
\begin{equation}
\frac{1}{i}\frac{d\psi}{dt}(t)+H\psi(t)=0,\quad
\psi(0)=\psi,\label{timesatisfiestimedependentSchrdoingerequation}
\end{equation}
\noindent which shows that $t$ introduced in Definition
\ref{localtime-def} exactly plays the role of time for the system
$(H,\HH)$ as it is assumed by equation
\eq{time-dependent-SchrodingerEquation}. However, the physical
role of $t$ as physical time is yet to be established.

Another nonstandard element implicit to Definition
\ref{localtime-def} follows from the fact, that the quantum
Universe as we currently perceive it consists of more than one
closed (``local'') system, each of which {\it independently}
satisfying the conditions of Definition \ref{localtime-def}. Hence
if the parameter $t$ in Definition \ref{localtime-def} plays the
role of physical time, the Universe consists of plenty of (at
least approximately) closed, i.e. local, systems, each of which
bearing its own local time generated by their respective local
Hamiltonians. To this end, a word of caution is in order. If we
regard the quantum Universe as the only truly closed quantum
system, i.e. if we do not allow for at least approximate
closed-ness of certain subsystems, we will not be able to describe
even a single act of quantum measurement within the unitary
quantum theory.

We strongly emphasize that the above introduction of time -- that
still requires a rigorous procedure of Theorem
\ref{asymptoticallyclassicalbehavior} below -- for a system
$(H,\HH)$ has been done with only using the notion of the {\it
time-independent} configuration and momentum operators $(x,p)$,
since time $t$ is defined solely through the use of a  local
system's Hamiltonian $H$ in \eq{HamiltonianofNbodysystem} which is
defined by $(x,p)$. In this sense, the notion of time is not, i.e.
not necessarily, any fundamental notion of universal importance
even if quantum mechanics can be formulated in time-dependent
fashion with using time-dependent Schr\"odinger equation as a
basic equation. Therefore a shift in the paradigm of Time [8]: (a)
we start from the time-less position and momentum operators, which
(b) define the time-independent Hamiltonian $H$, which generates
dynamics of a closed system in Definition \ref{localtime-def} and
(c) introduces time as an emergent property of the local system
with the link (d) one Hamiltonian, one local time for the local
(closed) system. In other words: the standard fundamental role of
the universal time is abandoned {\it due to} establishing the
fundamental role of the local system's dynamics, i.e. of the local
system's clock, Definition \ref{localtime-def}.

Hence the concept of local time provides a missing link in the
original Schr\" odinger's thoughts by introducing (local) time for
a closed quantum system in a consistent way. If the parameter $t$
may be regarded as a closed system's (local) time, then
eq.\eq{timesatisfiestimedependentSchrdoingerequation} is the
differential form of the universal fundamental dynamical law for
closed systems that for different Hamiltonians produces different
local times, i.e. different dynamics, such as those given by
eqs.\eq{(1)'}-\eq{(2)'}. Therefore there is no need to choose
between the dynamical equations \eq{(1)'} and \eq{(2)'}--they are
both correct for their respective local times. Needless to say,
eq.\eq{timesatisfiestimedependentSchrdoingerequation}
straightforwardly leads to derivation of the time-independent
equation \eq{stationarySchrodingerEquation}, which now becomes a
special case, i.e. non-fundamental physical law. However, bearing
in mind that, at its best, equation
\eq{timesatisfiestimedependentSchrdoingerequation} can serve as a
symptom of the physical nature of the parameter $t$ as the
physical time for local system, our argument requires the
following completion.

We now turn to the nature of local time which tells that the name
``time'' is appropriate for $t$. For simplicity we here consider
the two-body case $N=2$ only, whose proof is found in Lemma 5.2 in
\cite{KF}. For general $N\ge2$, see Theorem 1 in \cite{K}, Theorem
3.2 in \cite{KN}.
\begin{theorem}\label{asymptoticallyclassicalbehavior}
Let $\psi\in \HH_c(H)$ with $(1+|x|)^2\psi\in\HH=L^2(\R^3)$. Then
there is a sequence $t_m\to\infty$ ($m\to\infty$) such that for
any $\varphi\in C_0^\infty(\R)$ and $R>0$
\begin{eqnarray}
&&\Vert\chi_{\{x\in\R^3||x|<R\}}e^{-it_mH}\psi\Vert\to0,\\
&&\Vert(\varphi(H)-\varphi(H_0))e^{-it_mH}\psi\Vert\to0,\\
&&\left\Vert\left(\frac{x}{t_m}-\frac{p}{\mu}\right)e^{-it_mH}\psi\right\Vert\to0
\end{eqnarray}
as $m\to\infty$, where $p=-i\partial/\partial x$, $\mu$ is reduced
mass, and $\chi_B$ denotes the characteristic function of a set
$B$. The similar asymptotic relations hold for some sequence
$t_m\to-\infty$ $(m\to-\infty)$.
\end{theorem}
We note that $x$ denotes the distance operator at ``time'' $t$
from the origin around which the quantum particle is assumed to
have started at the initial ``time'' $t=0$. Thus the theorem tells
that for a scattering state $\psi$ belonging to the continuous
spectral subspace $\HH_c(H)$ for $H$, the local clock $e^{-itH}$
works such that the ``mechanical'' velocity $x/t_m$ becomes close
to quantum mechanical velocity $p/\mu=-\mu^{-1}i\partial/\partial
x$ as $m\to\infty$ on the state $e^{-it_mH}\psi$.
\begin{equation}
\frac{x}{t_m}\sim
\frac{p}{\mu}\quad(t_m\to\infty).\label{classicalquantumvelocities}
\end{equation}
This tells that the quantum mechanical wave function
$e^{-it_mH}\psi$ travels most densely around a trajectory of a
classical counterpart for which eq.\eq{classicalquantumvelocities}
would be equivalent with the classical time expressed in the
well-known form, $t=\mu x/p$, which, in turn, is sometimes used as
a basis of time quantization \cite{Y.AharonovD.Bohm}.

Hence the following answer to the first problem indicated in
Section 1: the $c$-number $t$ for a local clock $e^{-itH}$ assumes
the role of ``time" from classical mechanics.

\section{Fourier-Laplace transform of a local clock}\label{Fourier-Laplace}

Theorem 1 indicates that a closed system's Hamiltonian generates
dynamics, which, in turn, bears the system's local time. In this
section, we show that the inverse also holds, that is, we show
that local time $t$ established by Theorem 1 determines the
time-independent Hamiltonian of a closed system, that answers the
second problem indicated in Section 1.

First assume that $\HH_c(H)=\{0\}$. Then one has $\HH=\HH_p(H)$
and thus the space $\HH$ is spanned by just the eigenfunctions
$\psi$ for $H$. Hence the spectral property of $H$ is completely
determined by timeless Schr\"odinger equation
\eq{stationarySchrodingerEquation}. Appearance of time also for
this case regards the generic state $\psi=\sum_{j=1}^K a_j\psi_j$,
with $H\psi_j=E_j\psi_j$, $\psi_j\ne0$,  and thus
$|e^{-itH}\psi(x)|^2=\sum_{j,k=1}^Ke^{-it(E_j-E_k)}a_j\overline{a}_k\psi_j(x)\overline{\psi}_k(x)$
$\ne$ constant in general. This describes the operation of a local
clock on $\psi\in \HH_p(H)$. Therefore, what Schr\"odinger did in
\cite{{S1},{S2},{S3}} is to identify eigenvalues $E$ and
corresponding eigenfunctions $\psi$ of $H$. Thus his work in those
papers is an analysis of pure point spectrum $\sigma_p(H)$ of $H$
and the eigenspace $P(\lambda)\HH$ for $H$ with $\lambda\in
\sigma_p(H)$. This result clarified the structure of $H$ on the
eigenspace $\HH_p(H)\subset \HH$ of $H$. In other words, the
time-dependent analysis of $e^{-itH}$ on $\HH_p(H)$ is reduced to
the time-independent analysis of eigenvalues and the corresponding
eigenfunctions of $H$.

The inverse to this is as follows. It is known that the following
mean ergodic identity holds for each $\lambda\in\R$ (see Ch. 10 in
\cite{Kato}).
\begin{equation}
P(\lambda)=\mbox{s-}\lim_{t_2-t_1\to\infty}(t_2-t_1)^{-1}\int_{t_1}^{t_2}e^{-it\lambda}e^{itH}dt.
\end{equation}
Thus analysis of the time-independent Hamiltonian is reduced to
the analysis of the solution $e^{-itH}$ of time-de\-pen\-dent
 equation (4).

The analysis of other spectrum $\sigma_{ac}(H)=\sigma_c(H)$ of $H$
is reduced to the analysis of the absolutely continuous part
$H_{ac}$ of $H$, i.e. to the analysis of $H$ restricted to the
absolutely continuous subspace $\HH_{ac}(H)=\HH_c(H)$. As the
measure $(E_H(B)\psi,\psi)=\Vert E_H(B)\psi\Vert^2$ is absolutely
continuous for $\psi\in \HH_{ac}(H)$, there exists an integrable
differentiation $\frac{d}{d\lambda}(E(\lambda)\psi,\psi)$ for all
$\lambda\in\R$ such that for a Borel set $B$ of $\R$ the following
relation holds.
\begin{equation}
(E_H(B)\psi,\psi)=\int_B\frac{d}{d\lambda}(E(\lambda)\psi,\psi)d\lambda\quad(\psi\in\HH_{ac}(H)).
\end{equation}
\noindent Let $\tilde{\cal T}$ be the closed set of all
eigenvalues of $H$ and its subsystem Hamiltonians. Then it is
known (see Theorem 8.1 in \cite{PSS}) that for $\psi\in
L^2_\delta(\R^{3n})$ $(1\ge \delta>1/2)$ and $\lambda\in
\R\setminus \tilde{\cal T}$, the boundary value $R(\lambda\pm i0)$
as $\epsilon\downarrow 0$ of the resolvent $R(\lambda\pm
i\epsilon)$ exists as a bounded operator from a subspace
$L^2_\delta(\R^{3n})(\subset L^2(\R^{3n}))$ into its dual space
$L^2_{-\delta}(\R^{3n})$, and satisfies for $\psi\in
L^2_\delta(\R^{3n})$
\begin{equation}
\frac{dE}{d\lambda}(\lambda)\psi=\frac{1}{2\pi
i}(R(\lambda+i0)-R(\lambda-i0))\psi\in L_{-\delta}^2(\R^{3n}).
\end{equation}
In general, by \eq{resolutionofidentity} local clock
$e^{-itH}\psi$ for $\psi\in\HH_{ac}(H)$ is given by a Fourier
transform of $dE(\lambda)$ so that we have
\begin{equation}
\begin{aligned}
e^{-itH}\psi&=\int_{\R}e^{-it\lambda}dE(\lambda)\psi\\
&=\int_{\R}e^{-it\lambda}\frac{dE}{d\lambda}(\lambda)\psi d\lambda\\
&=\frac{1}{2\pi
i}\int_{\R}e^{-it\lambda}(R(\lambda+i0)-R(\lambda-i0))\psi
d\lambda.
\end{aligned}\label{e-itHpsi}
\end{equation}
This shows that the analysis of time evolution or local clock
$e^{-itH}\psi$ of the system with Hamiltonian $H$ for $\psi\in
\HH_{ac}(H)$ can be reduced to the analysis of the boundary values
of the resolvent $R(\lambda\pm i\epsilon)\psi$ as
$\epsilon\downarrow0$. Hence time-dependent analysis of quantum
mechanics on $\HH_{ac}(H)$ can be derived from time-independent
analysis of QM.

Conversely, writing $R(z)=(H-z)^{-1}=(H_0+V-z)^{-1}$ for
$z\in\C\setminus\R$, we have
\begin{equation}
\begin{aligned}
R(z)\psi &=(H-z)^{-1}\psi=\pm i\int_0^{\pm\infty}
e^{itz}e^{-itH}\psi dt \\ &(\pm\mbox{Im }z>0, \psi\in \HH).
\end{aligned}
\end{equation}
This shows that the analysis of the boundary values of the
resolvent $R(z)$ is reduced to the analysis of the convergence of
Fourier-Laplace transform of the local clock $e^{-itH}$ when
$|\mbox{Im }z|\to 0$. In this sense, the analysis of spectral
property of the time-independent Hamiltonian $H$ can be reduced to
the analysis of the solution of the time-dependent Schr\"odinger
equation.

These show that the time-dependent analysis and time-independent
analysis of QM are equivalent for $\psi\in\HH_{ac}(H)$. Together
with the result we have shown for $\psi\in \HH_p(H)$, these
provide the desired argument, which now can be stated as

\begin{theorem}\label{equivalence}
Time-dependent analysis and stationary analysis of Quantum
Mechanics are mutually equivalent for closed  systems.
\end{theorem}

\section{Quantum field theory}\label{QFT}

We consider in this section how the local time works in the case
of Quantum Field Theory (QFT). In QFT that ignores the spin of the
system, Hamiltonian of a system is given as follows. Let $q(x),
p(x)$ be maps from $\R^3$ into a space of selfadjoint operators in
a Hilbert space such that the following canonical commutation
relations hold for all $x,x'\in\R^3$.
\begin{equation}
\begin{aligned}
&[q(x),p(x')]=i\delta(x-x'),\\
&[q(x),q(x')]=[p(x),p(x')]=0.
\end{aligned}
\end{equation}
Then the Hamiltonian $H$ is defined by
\begin{equation}
H=\frac{1}{2}\int(p(x)^2+c^2\nabla
q(x)^2+c^4\mu^2q(x)^2)dx.\label{HamiltonianQFT}
\end{equation}
We assume that $H$ defines a selfadjoint operator in a suitable
Hilbert space. Then we can define local clock and time of the
system by the evolution $e^{-itH}$ as in Definition
\ref{localtime-def}. Using the local time of the system, we define
\begin{equation}\left\{
\begin{aligned}
&q(x,t)=e^{itH}q(x)e^{-itH},\\
&p(x,t)=e^{itH}p(x)e^{-itH}.
\end{aligned}\right.\label{evolutedqp}
\end{equation}
Let
$$
{\quad \quad \quad \ \ \ \ {\mbox{\it\scriptsize $j\atop\smile$}}
\ \  \ \ \  \ }\atop{a^{(j)}=(0,\dots,0,a,0,\dots,0)}
$$
be a vector with $j$-th component being $a$ and others zero. Then
recalling that
\begin{equation}
\begin{aligned}
&\nabla q(x)=\left(\frac{\partial q}{\partial{x_j}}\right)_{j=1}^3,\\
&\frac{\partial
q}{\partial{x_j}}(x)=\lim_{a\to0}\frac{q(x+a^{(j)})-q(x)}{a},
\end{aligned}
\end{equation}
we have
\begin{equation}
\begin{aligned}
&\left[ \frac{\partial
q}{\partial{x_j}}(x),p(x')\right]=i\frac{\partial\delta}{\partial{x_j}}(x-x'),\\
&[\nabla q(x),p(x')]=i\nabla \delta(x-x').
\end{aligned}
\end{equation}
\medskip

\noindent From this and \eq{HamiltonianQFT} follows:
\medskip

\noindent
\begin{theorem}\label{QFTfundamentalequation} For $(q,p)$ defined above, we have
\begin{equation}\left\{
\begin{aligned}
&\frac{\partial q}{\partial t}(x,t)=p(x,t),\\
&\frac{\partial p}{\partial t}(x,t)=(c^2\Delta q-c^4\mu^2q)(x,t).
\end{aligned}\right.
\end{equation}
Therefore we have
\begin{equation}
\left(\frac{1}{c^2}\frac{\partial^2}{\partial
t^2}-\Delta+c^2\mu^2\right) q(x,t)=0.\label{QFTwaveequation}
\end{equation}
This holds if the following holds
\begin{equation}
\frac{1}{i}\frac{\partial q}{\partial
t}(x,t)+c\sqrt{-\Delta+c^2\mu^2} q(x,t)=0.\label{qxt}
\end{equation}
\end{theorem}
If we take $c$ equal to the speed of light, the equation
\eq{QFTwaveequation} becomes covariant with respect to the Lorentz
transformations, and thus this equation successfully describes the
free relativistic field. The obtained equation
\eq{QFTwaveequation} is the Klein-Gordon equation and shows that
the field propagates as a wave. Furthermore \eq{qxt} shows that to
adopt Hamiltonian $H$ in \eq{HamiltonianQFT} is equivalent to
adopting the Hamiltonian
\begin{equation}
H_{(r)}=c\sqrt{-\Delta+c^2\mu^2}
\end{equation}
and the local clock $e^{-itH_{(r)}}$ for the same QF system. Thus
the corresponding Schr\"odinger equation
\begin{equation}
\frac{1}{i}\frac{dq}{dt}(t)+H_{(r)}q(t)=0.\label{FieldSchrodinger}
\end{equation}
is a fundamental equation for free relativistic quantum field
theory.

We recall that all this is formally given through the use of Fock
space $F=\bigoplus_{k=0}^\infty \HH^{n}$
$(\HH^n=\overbrace{\HH\otimes\cdots\otimes\HH}^{n\
\mbox{\scriptsize{factors}}}$)
 and a selfadjoint operator in $F$ like
\begin{equation}
H=\sum_{k=0}^\infty \omega_ka_k^\dagger a_k,
\end{equation}
where $\omega_k=c\sqrt{k^2+c^2m^2}$, and $a_k^\dagger$ and $a_k$
are creation and annihilation operators, respectively.

\section{Discussion}\label{Discussion}

Modern open quantum systems theory \cite{{BreuerPetruccione},
{RivasHuelga}} offers a unique physical basis for the explicit
appearance of time for non-conservative systems. Time dependence
of the system's Hamiltonian may be due to the environmental
influence, e.g. \cite{BriggssRostEPJD}. That is, explicit time
dependence in a quantum system's Hamiltonian may be a symptom of
the system's interaction with another system that is often called
environment. Hence whenever we start with arguments regarding open
systems, we may ultimately end up with a closed system
\cite{{BriggssRostEPJD}, {BreuerPetruccione}, {RivasHuelga}} that
is described by the fundamental dynamical law
\eq{time-dependent-SchrodingerEquation} and hence with the
conclusion that all subsystems (conservative or not) of a closed
system share the same physical time.

As we emphasized in Introduction, the standard global and
universal time common for all subsystems (degrees of freedom) of
the quantum Universe appears as an assumption additional to the
fundamental postulates of quantum theory. Bearing in mind
Definition 2, this assumption is, in our opinion, a huge step
requiring justification or otherwise becomes Procrustean.

On the other hand, the concept of emergent local time, Definition
2, neither relies nor it requires any  assumptions additional to
the postulate of the fundamental unitary dynamics in quantum
theory. Local Time paradigm \cite{K} establishes physical time for
a single closed system without any intrinsic inconsistencies,
which are otherwise found for some concurrent approaches to the
concept of local time (or 'multi time') in the non-relativistic
context \cite{PetratTumulka}.

Some details and ramifications regarding the concept of local time
can be found in Refs. \cite{{K}, {Ki-Fl}, {Dugic-PRSA1}} while
certain corollaries of Local Time paradigm can be found in
\cite{Dugic-PRSA1, {Dugic-PRSA2}}. Interpretational consequences
and links with the existing approaches to time in quantum theory
will be presented elsewhere.

\section{Conclusion}\label{Conclusion}

The concept of Local Time \cite{K} is a minimalist alternative to
the standard concept of universal time in the unitary quantum
theory. The unitary dynamics bears local time as an internal
characteristic that is neither fundamental nor universal on the
quantum-mechanical level while being consistently attributable  to
every, at least approximately, closed quantum system.

\label{lastpage-01}

\end{document}